\documentclass[12pt]{iopart}
\usepackage{graphicx,color}
\usepackage[switch]{lineno}
\def\a{\alpha}
\def\b{\beta}
\def\G{\Gamma}
\def\d{\delta}
\def\D{\Delta}
\def\e{\epsilon}
\def\z{\zeta}
\def\h{\eta}
\def\th{\theta}
\def\Th{\Theta}
\def\l{\lambda}
\def\La{\Lambda}
\def\p{\pi}
\def\r{\rho}

\def\ps{\psi}
\def\Ps{\Psi}
\def\o{\omega}
\def\O{\Omega}
\def\i{\int}
\def\bx{{\mathbf x}}
\def\bq{{\mathbf q}}
\def\bp{{\mathbf p}}
\def\bn{{\mathbf n}}
\def\Tr{\mbox{Tr}}
\def\be{\begin{equation}}
\def\ee#1{\label{#1}\end{equation}}

\begin{document}

\title{Statistics of work and fluctuation theorems for microcanonical initial states }

\author{Peter Talkner$^1$, Manuel Morillo$^2$, Juyeon Yi$^3$ and Peter H\"anggi$^1$ }
\address{$^1$ Institute of Physics, University of Augsburg, Universit\"{a}tsstrasse 1, D-86135 Augsburg, Germany}
\address{$^2$ Fisica Te{\'o}rica, Universidad de Sevilla, Apartado de Correos 1065, Sevilla 41080, Spain}
\address{$^3$ Department of Physics, Pusan National University, Busan 609-735, Korea}
\ead{peter.talkner@physik.uni-augsburg.de}

\date{\today}

\begin{abstract}
The work performed on a system in a microcanonical state by changes in a control parameter is characterized in terms of its  statistics.
The transition probabilities between eigenstates of the system Hamiltonians  at the beginning and the end of the parameter change obey a detailed balance-like relation from which various forms of the microcanonical fluctuation theorem are obtained. As an example, sudden deformations of a two dimensional harmonic oscillator potential are considered and the validity of the microcanonical Jarzynski equality connecting the degrees of degeneracy of energy eigenvalues before and after the control parameter change is confirmed.

\end{abstract}


\section{Introduction}
The microcanonical ensemble has always played an important role in the foundation of equilibrium statistical mechanics \cite{G,Gibbs,M,K,Campisi}. For practical applications and computational purposes though the canonical and the grandcanonical ensembles have traditionally been considered more relevant because
frequently the considered systems are in contact with their environments acting as heat and particle reservoirs. Often, the choice of the ensemble is merely a matter of convenience because
for the large class of systems with short range interactions all ensembles become equivalent provided the number of constituting particles is sufficiently large.

For systems with long range interactions \cite{BMR} and for small systems with few particles the different ensembles are no longer equivalent and hence the correct ensemble must be chosen depending on the physical situation \cite{CK}, -- with gravity being a prominent example \cite{Gross}. The enormous recent progress in manipulating cold atoms \cite{Ph}, which often are isolated from their environment to a very high degree, has given renewed interest and practical relevance to the microcanonical ensemble.

In experiments, such systems are tested by the change of control parameters such as the strengths and positioning of laser-fields which generate the effective potentials in which the atoms are trapped \cite{BDZ}.
The work performed on the system turns out to be a random quantity with a distribution depending both on the protocol and the chosen initial state \cite{CTH}. In the case of a microcanonical system  the work distribution depends on the initial energy.

The general form of the microcanonical work distribution was first derived for classical systems in \cite{CBK} and later for quantum systems in \cite{CTH,THM,CPRE} where a formal representation of a microcanonical state was used in terms of a Dirac delta-function.
In order to obtain a properly defined density matrix, i.e., a positive operator with unit trace  \cite{H}, the delta function defining it must be regularized. This regularization can be done in many different ways and therefore introduces some arbitrariness. It can physically be understood as the attempt
to account for experimental or theoretical uncertainties of the energy.

In the present work we take a different approach and write the density matrix of a microcanonical state as the properly normalized projection operator onto all states with a given energy as discussed in Section II. Based on this form of the density matrix we derive the probability density function of work and various equivalent forms of the microcanonical fluctuation theorems in Section III.  These fluctuation theorems are a consequence of a detailed balance-like relation for the transition probabilities between energy eigenstates of the Hamiltonians at the end and the beginning of the protocol, a relation that is also crucial for canonical \cite{J,C,Tasaki,TH} and grandcanonical fluctuation theorems \cite{YKT}. We illustrate the obtained results with the example of a two-dimensional harmonic oscillator whose potential energy is suddenly squeezed or widened along its principal axes. Work statistics for a one-dimensional oscillator under the action of time-dependent but otherwise constant forces was analyzed for different initial conditions including pure energy eigenstates in \cite{TBH}. The squeezing of a one-dimensional oscillator out of canonical states was studied in \cite{DL}.
Here we consider rational principal frequency ratios before and after the deformation giving rise to degeneracies of the energy eigenvalues and compare the quantum work distributions to the corresponding classical expressions which are typically approached for large initial energies. The paper closes with concluding remarks. In the appendix, the work distributions  for deformations of classical one- and two-dimensional oscillators are presented.

\section{Microcanonical state}
In order to prevent a finite system to eventually escape to infinity, it has to be confined in space by a sufficiently steeply increasing potential. As a consequence, the Hamiltonian of such a system has a pure point-spectrum. Accumulation points and continuous parts of the spectrum do not occur.\footnote{A mathematically rigorous formulation of conditions leading to many-particle Hamiltonians with discrete spectra can be found in \cite{BR}.}     

There are various ways to represent the density matrix of a system in the microcanonical state. One, formal representation is most close to the classical form of a microcanonical probability density function (pdf) in terms of a Dirac-delta function concentrated on the respective energy shell. It is given by
\be
\r_E=\o_E^{-1} \d(H-E)\:,
\ee{rd}
where $H$ is the Hamiltonian of the considered system and
\be
\o_E = \Tr \d(H-E)
\ee{dos}
denotes the density of states. In order that the density matrix as well as the density of states are well defined for all values of the continuous energy variable $E$ the delta-function $\d(x)$ must be replaced by a continuous function which is mainly localized at $x=0$, such as a narrow Gaussian $\d_\e(x)$,
\be
\d_\e(x) = 1/\sqrt{2 \p \e^2} e^{-x^2/(2 \e^2)}\:.
\ee{de}
The width $\e$ has to be chosen small enough. It may be thought of representing the actual resolution of the energy as being determined by experiment or, within a theoretical model.

With the spectral representation of the Hamiltonian,
\be
H = \sum_n E_n P_n\:,
\ee{H}
expressed in terms of the eigenenergies $E_n$ and the eigenprojectors $P_n$ the density matrix becomes a sum over the energy levels $n$, i.e.
\be
\r_E = \o_E^{-1} \sum \d(E_n-E) P_n\:.
\ee{rs}
The density of states then is a weighted sum of delta-functions
\be
\o_E = \sum_n d_n \d(E_n-E)\:,
\ee{doss}
where $d_n$ denotes the degeneracy of the $n$th eigenvalue, expressible as the trace of the respective projector, $d_n = \Tr P_n$.

Alternatively, the microcanonical density matrix can be written in terms of an eigenprojector
\be
\r_n = d_n^{-1} P_n\:,
\ee{rn}
which is a function of the discrete index $n$ rather than the continuous energy $E$. The Dunford integral \cite{DS} allows one to define an operator $P(E)$ that coincides with $P_n$ for $E=E_n$ and is zero otherwise.
It is defined in terms of an operator-valued Cauchy-integral reading
\be
P(E) = \oint_{\mathcal{C}_\e(E)} \frac{dz}{2 \p i}\frac{1}{z-H}\:,
\ee{PDS}
where $\mathcal{C_\e(E)}$ is a circular path with radius $\e$ encircling $E$ counterclockwise.
The microcanonical density matrix as a function of $E$ can then be represented as
\be
\r(E) = \mathcal{N}_E P(E) \:,
\ee{rE}
where
\be
\mathcal{N}_E = \sum_n d^{-1}_n \z(E_n-E)\:,
\ee{NE}
and
\be
\z(x) = \left \{
\begin{array}{ll}
1 \quad&\mbox{for}\; x=0\\
0& \mbox{else}
\end{array}
\right .\:.
\ee{z}
In terms of the latter function, the projection operator $P(E)$ can also be expressed as
\be
P(E) = \sum_n \z(E-E_n) P_n\:.
\ee{PEPn}
This result follows from (\ref{PDS}) by means of the spectral resolution of the Hamiltonian $H$, (\ref{H}), and of the identity $\lim_{\e \to 0}\oint_{\mathcal{C}_\e(E)} dz/(z-E_n) = 2 \p i \z(E-E_n)$.
Note that, strictly speaking, the density matrices for the microcanonical state as given by the regularized delta-function representation $\r_E$,  see  (\ref{rd},\ref{de}), are different from $\r(E)$ defined by  (\ref{rE}). While $\r(E)$ is strictly zero for energies not agreeing with an eigenvalue $E_n$, $\r_E$ assigns a properly normalized density matrix to any value of the energy. For values different from an energy eigenvalue, the assigned states are fictitious.  Taking the idea of a microcanonical  ensemble  in a wider sense, namely that of  a collection of nearly identical systems with almost identical energy spectra within the small energy width $\epsilon$, the use of  $\r_E$ as a microcanonical density operator becomes, however,  physically sensible again.

\section{Work statistics and fluctuation theorem}
In order to perform work on a system, parameters $\l$ of the system's Hamiltonian are changed according to a prescribed protocol $\La =\{\l(t)|t_0 \leq t \leq t_f \}$ starting at $t_0$ and ending at time $t_f$. The ensuing dynamics is
governed by the Schr\"odinger equation
\begin{eqnarray}
i \hbar \partial U_{t,s}/\partial t &= H(\l(t)) U_{t,s}\:,\nonumber \\
U_{s,s} &= 1
\label{Uts}
\end{eqnarray}
for the unitary time evolution operator $U_{s,t}$,  where $H(\l(t))$ denotes the Hamiltonian with the parameter $\l(t)$ at the time $t$ according to the protocol.
Applied to the full duration of the protocol $\La$ the time evolution operator $U(\La)\equiv U_{t_f,t_0}$ describes the action of the protocol on pure states of the system.

As a unitary map $U(\La)$ can always be reversed by formally letting time  $t$ run backwards. If the parameters $\l(t)$ have a parity $\e_\l$ and if the time reversal operator $\th$ transforms the Hamiltonian $H(\l(t))$ at each time according to
\be
\th H(\l(t))\th^{-1} = H(\e_\l \l(t))
\ee{ThHTh}
then the inverse of the map corresponds to a dynamics in physical time under the action of the
time-reversed protocol
$\bar{\La} = \{\e_\l \l(t_f+t_0-t)|t_0\leq t \leq t_f \}$ 
\cite{CTH,AG,CTH2011}, i.e.
\be
U^{-1}(\La) = U^{\dagger}(\La) = \th U(\bar{\La}) \th^{-1}\:.
\ee{Utr}
For the sake of simplicity, so far, we have neglected a possible dependence of the Hamiltonian on parameters that are odd under time-reversal but unchanged during the protocol, as for example a constant magnetic field. Such fields  couple to system operators that are odd under time reversal and consequently also must be reversed in the time-reversed map   $\th U(\bar{\La}) \th^{-1}$.

In order to determine the work applied to the system in each individual run of the protocol, energy measurements at the beginning and the end of the protocol have to be performed. We assume here that the time-dependent Hamiltonian is gauged in such a way that it represents the instantaneous energy of the system at any instant of time \cite{CTH}. Assuming an initially uniform distribution of states with the same energy $E_n(t_0)$ the protocol $\La$ causes a transition to a state with final energy $E_m(t_f)$ with the probability $P_\La(m|n)$ given by
\be
P_\La(m|n) = d^{-1}_n(t_0)\Tr P_m(t_f) U(\La) P_n(t_0) U^{\dagger}(\La)\:,
\ee{Pmn}
where any $P_n(t)$ projects on the instantaneous eigenspace of the Hamiltonian $H(\l(t))$ corresponding to the instantaneous eigenenergy $E_n(t)$, i.e.,
\be
H(\l(t)) P_n(t) = E_n(t) P_n(t)\:.
\ee{HPE}
These projection operators are orthogonal ($P_n^{\dagger}(t) =P_n(t),\; P_n(t)P_m(t)= \d_{n,m}$), and complete ($\sum_n P_n(t) =1$); the degeneracy of the respective eigenvalue is determined by the trace $d_n(t) =\Tr P_n(t)$.
As an immediate consequence of the time-reversal relation (\ref{Utr}) one finds the detailed balance-like relation
\be
P_\La(m|n) d_n(t_0) = P_{\bar{\La}}(n|m) d_m(t_f)\:,
\ee{db}
where $P_{\bar{\La}}(n|m)$ denotes the transition for the backward protocol.
Summing both sides of this equation over $n$, one obtains the relation
\be
d_m(t_f) = \sum_n P_\La(m|n) d_n(t_0)
\ee{dndm}
expressing the degeneracy indices $d_m(t_f)$ at the end of the protocol with those at the beginning. In this way, also the microcanonical entropy at the end of the protocol is related to the entropy at the beginning. Here, the entropy can be calculated on the basis of Gibbs' definition in terms of the total number $D_E(t) = \i^E dE' \o_{E'}=\sum_{n:E\geq E_n(t)} d_n(t) $ of states below the energy $E$, i.e. as $S_E(t) = k_B \ln D_E(t)$  \cite{G,Gibbs,M,CK,Hertz,Dunkel}. As a consequence of (\ref{dndm}), the total number $D_{E_n}(t_f)$ at the end of the protocol is related to the respective quantity at the beginning in the following way
\be
D_{E_n}(t_f) = \sum_{n'\leq n} \sum_m (P_\La(n'|m) - P_\La(n'|m+1)) D_{E_m}(t_0)\:.
\ee{DD}

A detailed balance-like relation, (\ref{db}), also holds for the energy-transition probability
\be
P_\La(E|E') = \sum_{n,m} \delta(E-E_n(t_f)) \z(E'-E_m(t_0))P_\La(n|m)\:,
\ee{PEE}
reading
\be
P_\La(E|E') \o_{E'}(t_0) = P_{\bar{\La}}(E'|E) \o_E(t_f)\:.
\ee{PP}
Using (\ref{dndm}) and the identity
\be
\sum_k \z(E-E_n)\d(E-E_k) = \d(E-E_n)
\ee{zd}
one proves  (\ref{PP}) by inspection.
\subsection{Work pdf and fluctuation theorems}
Since the work performed on the system during the protocol is determined as the difference of the final and initial energies, $w=E_m(t_f)-E_n(t_0)$, the pdf of work, $p_\La(w|n)$, conditioned on the discrete energy $E_n(t_0)$ is given by
\be
p_\La(w|n) = \sum_m \d(w-E_m(t_f)-E_n(t_0)) P_\La(m|n)\:.
\ee{pLan}
With (\ref{db}) one obtains a fluctuation theorem of the form
\be
p_\La(E_m(t_f)-E_n(t_0)|n)\:d_n(t_0) = p_{\bar{\La}}(E_n(t_0) - E_m(t_f)|m)\: d_m(t_f).
\ee{pmndn}
Alternatively, the work pdf conditioned on the continuous energy $E$ can be written as
\begin{eqnarray}
p_\La(w|E)& = \i dE' \d(w-E'+E) P_\La(E'|E) \nonumber \\
&= P_\La(E+w|E)\:.
\label{pLawE}
\end{eqnarray}

With  (\ref{PP}) we recover the  Tasaki-Crooks fluctuation theorem  \cite{CTH,C,Tasaki,TH}, generalized here for the microcanonical quantum case; i.e.,
\be
p_\La(w|E) \o_E(t_0) =p_{\bar{\La}}(-w|E+w) \o_{E+w}(t_f)\:.
\ee{FT}
This relation identically holds  for classical systems \cite{CBK}. For quantum systems it was derived in \cite{THM} by use of the delta-function representation of the initial microcanonical state.\footnote{Likewise, a similar quantum fluctuation theorem holds which involves the integrated density of states, relating to the Gibbs entropy, if the system is prepared with the properly normalized density operator due to Ruelle involving the step-function $\Theta [E - H(\lambda(t_0))]$, see relation (26) in \cite{THM}.}

In contrast to the classical case, the work for confined quantum systems with a finite number of degrees of freedom is always discretely distributed. Hence, the work pdf takes the form
\be
p_\La(w|E) = \sum_m \d(w-w_m) q_\La(w_m|E)\:,
\ee{pq}
where
\be
q_\La(w_m|E) = \i dw \z(w-w_m) P_\La(E+w|E).
\ee{qP}
For the discrete work probabilities $q_\La(w|E)$ the fluctuation theorem (\ref{FT}) takes the following, equivalent form:
\be
q_\La(w_m|E) \o_E(t_0) = q_{\bar{\La}}(-w_m|E+w_m) \o_{E+w_m}(t_f)\:.
\ee{ftqq}

The work pdf $p_\La(w;\b)$ for a canonical initial state can be determined from the microcanonical pdf by a Laplace transformation weighted by the density of states, \cite{THM}, i.e.,
\be
p_\La(w;\b) = Z^{-1}_\b(t_0) \i dE \o_E(t_0) e^{-\b E} p_\La(w|E)\:,
\ee{pbE}
where $\b = 1/(k_B T)$ is the inverse temperature of the initial canonical state and where the partition function $Z_\b(t)$ is given by the Laplace transform of the density of states yielding
\be
Z_\b(t) = \i dE \o_E(t) e^{-\b E}\:.
\ee{Zb}
Applying the Laplace transform on both sides of the microcanonical fluctuation theorem (\ref{FT}) immediately gives the well-known canonical  Tasaki-Crooks relation \cite{Tasaki,TH}
\be
Z_\b(t_0) p_\La(w;\b) = e^{\b w} Z_\b(t_f) p_{\bar{\La}}(-w;\b)\:.
\ee{cft}

\subsection{Characteristic function}
The characteristic function defined as the Fourier transform of the work pdf  \cite{TLH},
\be
G_\La(u|E) = \i dw e^{i u w} p_\La(w|E)
\ee{Gup}
can be expressed in terms of a correlation function reading \cite{THM}
\be
G_\La(u|E) = \Tr e^{iu H^H(t_f)} e^{-i u H(t_0)} \r_E\:,
\ee{Gur}
where
\be
H^H(t_f) = U^\dagger(\La) H(t_f) U(\La)
\ee{HH}
denotes the Hamiltonian in the Heisenberg picture at the end of the protocol.

In the canonical case, the characteristic functions of the forward and the backward processes, $G_\La(u;\b)$ and $G_{\bar{\La}}(u;\b)$, defined in terms of the Fourier transforms of the respective canonical work pdfs, are related by  the fluctuation theorem reading \cite{TLH}
\be
G_\La(u;\b) = G_{\bar{\La}}(-u+i\b;\b)
\ee{GG}
which is equivalent to the canonical Tasaki-Crooks relation (\ref{cft}). An analogous relation for the characteristic functions does  {\it not} exist in the case of a microcanonical initial condition. This is due to the fact that in the microcanonical Tasaki-Crooks relation the initial conditions for the forward and the backward processes not only differ with respect to the parameter values but also with respect to the initial energies, $E$ and $E+w$, in contrast to the canonical case where the temperatures of the initial and final states are equal.

For the characteristic function
\be
G_\La(u|n) = \i dw e^{i u w} p_\La(w|n)
\ee{Gwn}
the fluctuation theorem (\ref{pmndn}) implies the sum-rule
\be
\sum_n G_\La(u|n) d_n(t_0) = \sum_n G_{\bar{\La}}(-u|n) d_n(t_f)
\ee{sGn}
which is meaningful for systems with a finite dimensional Hilbert space; for infinite-dimensional Hilbert spaces the sums need not converge. Likewise one obtains a relation for the characteristic function depending on the energy $E$, (\ref{Gup}), reading
\be
\i dE G_{\La}(u|E) \o_E(t_0) = \i dE G_{\bar{\La}}(-u|E) \o_E(t_f)\:,
\ee{iGE}
provided the Hilbert space of the system is finite.
\section{Quench of a 2d harmonic oscillator}
As an illustrative example we consider the work performed on a particle of mass $m$ upon a sudden change of the principal curvatures of a harmonic potential, described by
\be
U_\a(\bx) =\frac{m}{2} \left [(\o_1^{(\a)})^2 x^2_1 + (\o_2^{(\a)})^2 x^2_2  \right ]\:,
\ee{Ua}
where $\a=0$ refers to the potential before and $\a=f$ to the potential after the quench.
Accordingly, the Hamiltonians before and after the quench read
\be
H_\a=\frac{\bp^2}{2m} + U_\a(\bx)\:,
\ee{Ha}
where $\bp =(p_1,p_2)$ denote the momenta and $\bx =(x_1,x_2)$ the positions of the particle; their components relate to the principal components of the Hamiltonian.
The eigenvalues and the eigenvectors follow from those of the principal components and, hence, can be expressed as
\be
E^{(\a)}_{\bn^{\a}} = \hbar\left [(n^\a_1+1/2) \o^{(\a)}_1 + (n^\a_2+1/2) \o^{(\a)}_2 \right ] \:,\quad n^\a_i=0,1,2,\ldots
\ee{Enm}
and
\be
\Ps^{(\a)}_{\bn^{\a}}(\bx) =\ps^{(\a)}_{n^\a_1}(x_1) \ps^{(\a)}_{n^\a_2}(x_2)
\ee{Psps}
with the well-known 1d harmonic oscillator eigenfunctions
\be
\ps^{(\a)}_{n^\a_i}(x_i) = \sqrt{\frac{1}{2^{n^\a_i}\: n^\a_i! \:l^\a_i\: \sqrt{\p}}} \:H_n(\frac{x}{l^{\a}_i})\: e^{-x_i^2/(2 (l^\a_i)^2)}\quad i=1,2\:,
\ee{psn}
where $H_n(x)$ are the Hermite polynomials \cite{GR}, and $l^\a_i = \sqrt{\hbar/(m \o^{(\a)}_i)}$ is the characteristic length scale of the respective oscillator.
There are degenerate eigenvalues if the principal component frequencies $\o^{(\a)}_1$ and  $\o^{(\a)}_2$ stay in a rational ratio to each other, i.e.
if there are integer numbers $N^\a_i$ with
\be
\o^{\a}_i = N^\a_i \o^\a\:.
\ee{oNo}
For all rational cases, a coprime pair $N^\a_1,N^\a_2$, as well as a corresponding fundamental frequency $\o^\a$ are uniquely defined.
The eigenenergies can then be expressed in terms of an energy quantum number $K$ and the fundamental frequency reading
\be
E^{\a}_K =K \hbar \o^\a+ \frac{1}{2} \hbar \o^\a (N^\a_1 +N^\a_2)\quad K=0,1,2,\ldots
\ee{EK}
All principal component quantum numbers $n_1,n_2$
which are solutions of the linear Diophantine equation
\be
K=n_1 N^\a_1 + n_2 N^\a_2
\ee{KNn}
contribute to this energy.
The number of different pairs of non-negative integers solving this equation determines the degeneracy $d_K$ of the eigenenergy $E^\a_K$.

For a sudden change of the potential, the time-evolution map connecting the state of the system before and after the switch is unity, $U(\La) =1$, and, hence, the transition probability $p(\bn^{f}|\bn^{0})$ from the state $\Ps^{(0)}_{\bn^0}(\bx)$ to $\Ps^{(f)}_{\bn^f}(\bx)$ is given by the absolute square of the overlap integral of the two states, i.e. it becomes
\begin{eqnarray}
 p(\bn^{f}|\bn^{0})& = \left |\i d^2 \bx  \Ps^{(f)*}_{\bn^f}(\bx) \Ps^{(0)}_{\bn^0}(\bx) \right |^2 \nonumber \\
&= \left |\i dx_1 \ps^{(f)*}_{n^f_1}(x_1)\ps^{(0)}_{n^0_1}(x_1)\right |^2 \left |\i dx_2 \ps^{(f)*}_{n^f_2}(x_1)\ps^{(0)}_{n^0_2}(x_2) \right |^2\:.
\label{pn0nf}
\end{eqnarray}
A compact form of the absolute squares of the scalar products of the one-dimensional oscillator eigenfunctions with different frequencies was obtained in \cite{DAL} in terms of hypergeometric functions $_2F_1(a,b;c;z)$ \cite{GR}
\be
\fl
\left |\i dx \ps^{(f)*}_{n^f}(x)\ps^{(0)}_{n^0}(x) \right|^2 =
\left \{
\begin{array}{ll}
\displaystyle \frac{2 \l}{\l^2+1} \left (\frac{|\l^2-1|}{\l^2 +1} \right )^{(m+n)/2}&\\
\displaystyle \times \frac{\G(\frac{m+1}{2}) \G(\frac{n+1}{2})}{\p \G(\frac{1+m}{2}) \G(\frac{1+n}{2})}&\hspace{-1mm}\mbox{if}\;m,n\;\mbox{even}\\
\displaystyle \times\:  _2F_1(-\frac{m}{2},-\frac{n}{2};\frac{1}{2};\frac{4 \l^2}{(\l^2+1)^2})&\\*[6mm]
\displaystyle \frac{2^5 \l^3}{(\l^2+1)^3} \left (\frac{|\l^2-1|}{\l^2+1} \right )^{(n+m)/2-1} &\\
\displaystyle \times \frac{\G(\frac{n}{2}+1)\G(\frac{n}{2}+1)}{\p \G(\frac{n+1}{2}) \G(\frac{m+1}{2})} & \hspace{-1mm}\mbox{if}\;m,n\;\mbox{odd}\\
\displaystyle \times\: _2F_1(-\frac{m-1}{2},-\frac{n-1}{2};\frac{3}{2};\frac{4 \l^2}{(\l^2+1)^2})&\\*[6mm]
\displaystyle 0&\hspace{-1mm}\mbox{else}\:.\\
\end{array}
\right .
\ee{hyp}
Note that this integral only depends on the ratio of the lengths-scales $\l = l^f/l^0 = \sqrt{\o^0/\o^f}$ but not on the individual frequencies.

For irrational frequency ratios $\o^\a_1/\o^\a_2$ (\ref{pn0nf}) also yields the transition probability from energy $E^0_{\bn^0}$ to  $E^f_{\bn^f}$. In the case of frequency ratios which are rational both before and after the sudden curvature change, transitions between all states with prescribed initial and final energies contribute and hence one gets
\be
P(K^f|K^0) = d^{-1}_{K^0}(t_0)  \mathop{\sum_{\bn^0 \in K^0}}_{\bn^f \in K^f} p(\bn^f|\bn^0)\:,
\ee{PKfK0}
where we introduced the short-hand notation $\bn \in K$ for all solutions of  (\ref{KNn}). In all cases in which either of the two sets $\bn^\a \in K^\a$ is empty, the transition probability vanishes.
The forms of the energy transition probabilities in the remaining cases, in which the frequency ratio in the initial state is irrational and the final one irrational, or vice versa, are obvious and will not be considered here.

A sudden potential change starting from energy $E=\hbar \o^0 (K^0 +(N^0_1+N^0_2)/2)$ performs the work
\be
\fl
w=\hbar\left ( \o^f \left (K^f + (N^f_1+N^f_2)/2 \right) -\o^0\left (K^0 +(N^0_1+N^0_2)/2 \right ) \right )\quad K^f=0,1,2,\ldots
\ee{w}
with probability $q_\La(w|E) =P(K^f|K^0)$.
The minimal difference between two possibly occurring work values is
\be
\D w = \hbar \o^f\:.
\ee{Dw}
\begin{figure}\begin{center}
\includegraphics[width=8cm]{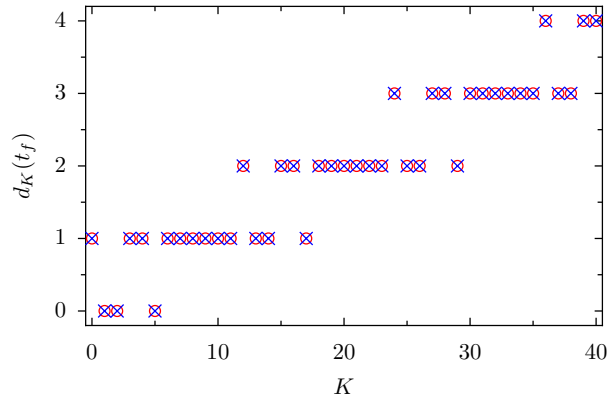}\end{center}
\caption{The degeneracy $d_K$ of an anisotropic oscillator with frequency ratio $4:3$ for different energy quantum numbers $K$ (red circles) is compared to the result of the fluctuation theorem (\ref{dndm}) (blue crosses) for a sudden change of the potential (\ref{Ua}) with $\o^0_1=2 \o$, $\o^0_2=5 \o$, $\o^f_1=4 \o$ and $\o^f_2=3 \o$. The agreement is perfect. The energies $E_K = (K+7/2) \hbar \o$ are eigenvalues of the post-quench Hamiltonian for all integers $K \geq 0$ apart from $K=1,2,5$. For these integers the respective equations (\ref{KNn}) reading $K=4 n_1 +3n_2$ do not have integer positive solutions and hence formally give  $d_K=0$. }
\label{fth}
\end{figure}

In figure~\ref{fth} we illustrate the validity of the fluctuation relation (\ref{dndm}) for a sudden protocol switching from a potential with fundamental frequency $\o^0$ and integer coefficients $N^0_1=2$, $N^0_2=5$ to the same fundamental frequency $\o^f = \o^0$, but with coefficients $N^f_1=4$, $N^f_2=3$.

\begin{figure}
\includegraphics[width=1\textwidth]{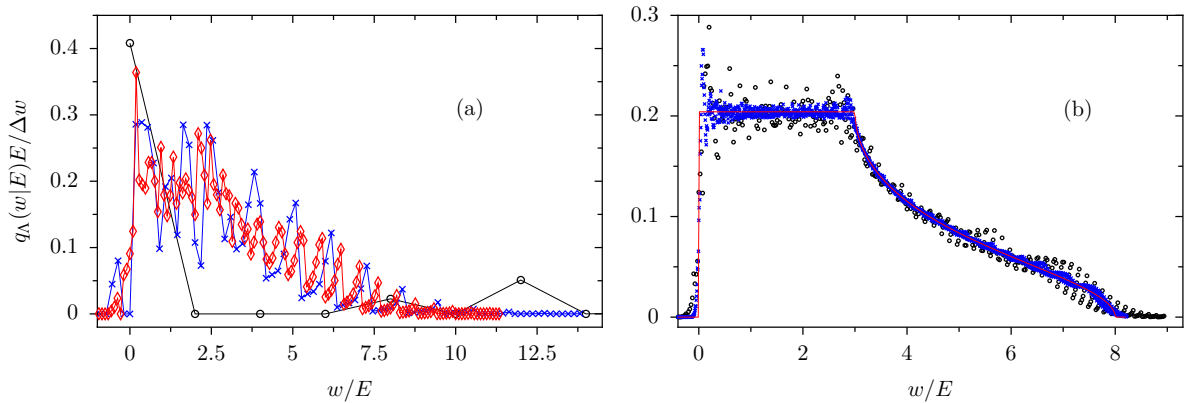}
\caption{The scaled work distribution $q_\La(w|E) E/\D w E$ is displayed as a function of $w/E$ for an initially isotropic oscillator, i.e. for  $N^0_1=N^0_2=1$, upon a sudden, anisotropic contraction of the potential characterized by $\o^f=\o^0$, $N^f_1=2$, $N^f_2=3$ at different initial energies $E= \hbar \o^0(L+1)$. The minimal distance between adjacent work values is $\D w = \hbar \o^f$. The panel (a) contains results for the low-lying initial energies with $L=0$ (black circles), $L=5$ (blue crosses) and $L=10$ (red diamonds); the thin lines are meant as guides for the eye. The panel (b) illustrates the convergence of the scaled work distribution towards the classical result (thin red line) given by the scaling function (\ref{r2}) with $\h_1=3$ and $\h_2=8$ for $L=40$ (black circles) and $L=160$ (blue crosses). With small but finite probabilities also values of the scaled work variable occur which fall outside of the classically allowed regime determined by the support $[0,8]$ of the scaling function $\r(z)$. Since $\h_1$ and $\h_2$  are both positive the scaling function is constant from $0$ up to $\h_1$ where its derivative is singular. }
\label{f4}
\end{figure}

\begin{figure}
\includegraphics[width=1\textwidth]{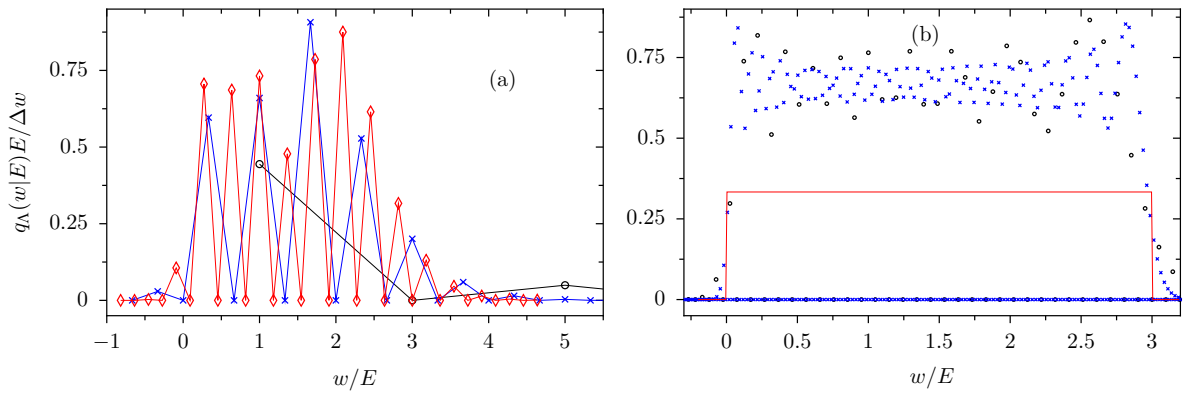}
\caption{The scaled work distribution resulting from a uniform contraction of an isotropic oscillator with $\o^f=\o^0$, $N^0_1=N^0_2=1$  and $N^f_1=N^f_2=2$ for different values of the initial energy $E= \hbar \o^0(L+1)$ with $L=0,5,10$ is displayed in the panel (a) and for $L=40,160$ as well as the classical scaling function  (\ref{r3}) in the panel (b). Symbols and color code are the same as for figure~\ref{f4}. In the present case of equal deformation parameters $\h\equiv \h_1=\h_2=3$ the scaling function $\r(w/E)$ is uniform on the interval $[0,\h]$.
Independently of the initial energy every other work value is forbidden because of strict selection rules. As a consequence, the scaled work distribution does not strictly converge to the classical scaling function for large values of the initial energy. Since the probability of half of the work values vanishes the probabilities of the allowed ones converge to twice the value given by the classical scaling function. Convergence to the classical limit is recovered if one restricts the statistics to the allowed work values and replaces $\D w = \hbar \o^f$ by $2 \hbar \o^f$ in the scaling of the work distribution.  }
\label{f6}
\end{figure}

\begin{figure}
\includegraphics[width=1\textwidth]{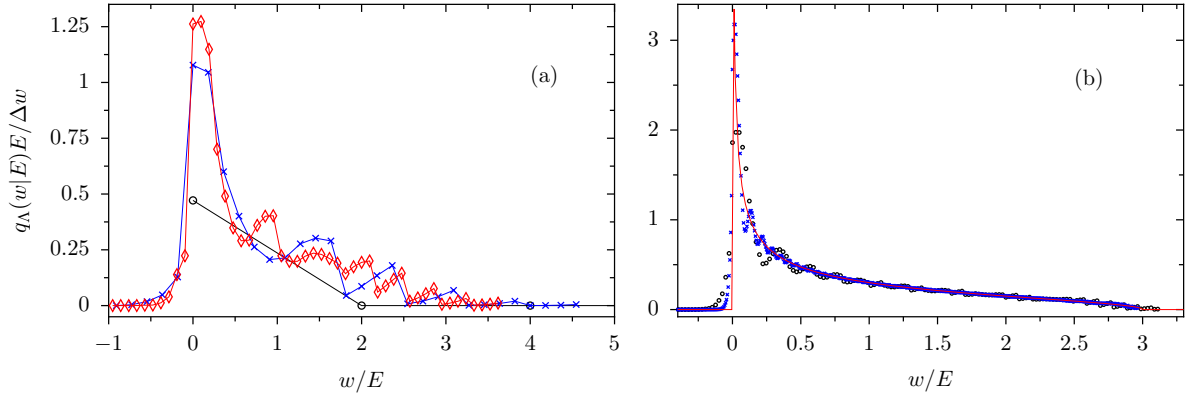}
\caption{The scaled work distribution that results from the squeezing of an isotropic harmonic potential in a single direction ($\o^0=\o^f$, $N^0_1=N^0_2=1$, $N^f_1=1$, $N^f_2=2$) is displayed for different initial energies  $E=\hbar \o^0(L+1)$ for $L=1,5,10$ in the panel (a) and $L=40,160$ together with the classical scaling function in the panel (b). Symbols and color code are chosen as in figure~\ref{f4}.
For large values of the initial energy the scaled distribution converges to the classical scaling function (\ref{r3}) for $\h_1=0$ and $\h_2=3$ with its square-root singularity at $w=0$.}
\label{f2}
\end{figure}
\begin{figure}
\includegraphics[width=1\textwidth]{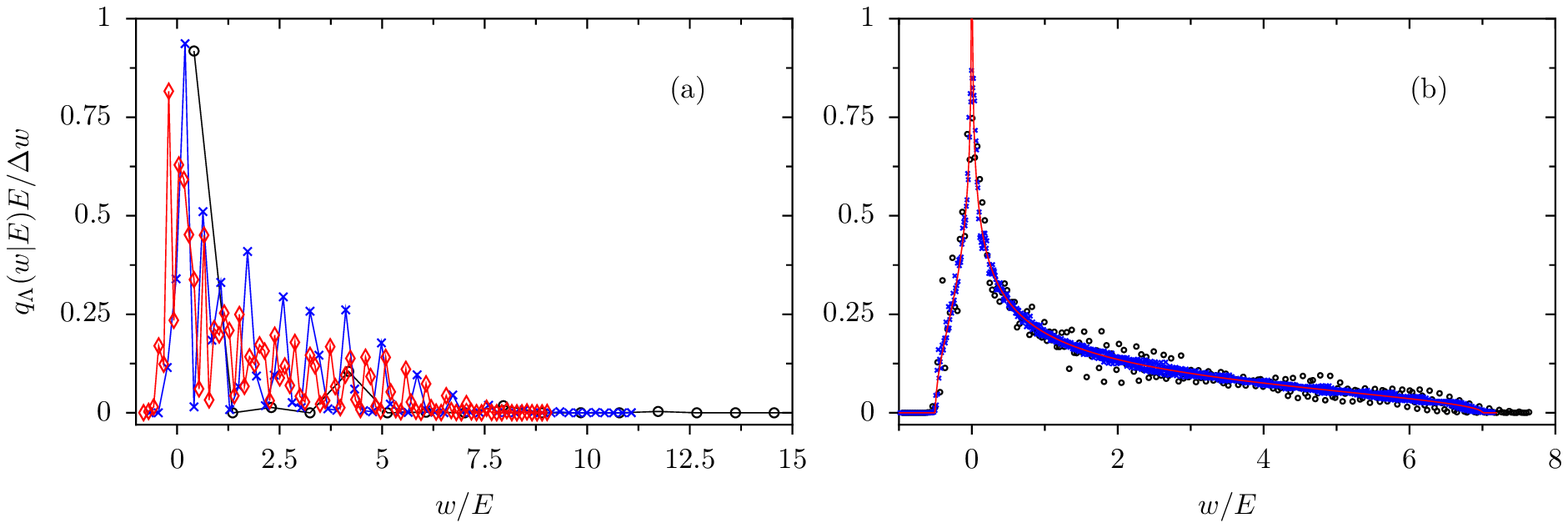}
\caption{The scaled work distribution for a simultaneous contraction and expansion of a harmonic potential in two orthogonal directions is maximal for work values close to zero. The ratio of the fundamental frequencies is $\o^f/\o^0=\sqrt{2}$; the  integers  before the deformation are $N^0_1=1$ and $N^0_2=2$ and after the deformation $N^f_1=2$ and $N^f_2=1$. Different initial energies $E= \hbar(L+3/2)$ are displayed for $ =1,5,10$ in the panel (a) and $L=40,160$ together with the classical scaling function in the panel (b). Symbols and color code are chosen as in figure~\ref{f4}. The classical scaling function with $\h_1=-1/2$ and $\h_2=7$ is approached for large initial energies. It has a logarithmic singularity at $w=0$. In the classical limit, the probability that the oscillator releases energy in this process is $p_+=\i_{\h_1}^0 \r(z)\approx 0.166$. }
\label{f8}
\end{figure}
In figures \ref{f4}-\ref{f8} the scaled probabilities $q_\La(w_n|E) E/\D w$ are displayed as functions of the scaled work $w/E$ for different sudden protocols and different initial energies and compared to the classical behavior.
The probability  $q_\La(w|E)$ is divided by the minimal work difference $\D w$ to make it comparable to a pdf, and the scaling of the work with the initial energy is motivated by the behavior of the work pdf $p^{\mbox{cl}}_\La (w|E)$ of a suddenly deformed classical
two-dimensional harmonic oscillator.
This pdf  scales with the initial energy according to
\be
p^{\mbox{cl}}_\La (w|E)= \r(w/E)/E\:,
\ee{pcl}
where the dependence of the scaling function $\r(x)$ on the protocol is completely determined by the dimensionless deformation parameters $\h_i$ defined as
\be
\h_i=\left (\o^f_i/\o^0_i \right )^2-1\:.
\ee{hi}
The deformation parameters  are restricted to  $-1<\h_i<\infty$; positive values correspond to compressions, and negative values to extensions of the potential.
The scaling function behaves qualitatively different depending on whether the product $\h_1 \h_2$ is positive or negative, i.e. whether both principal directions undergo the same types of deformation or opposite ones.
For further details and explicit expressions of the scaling function see the appendix.

Figure~\ref{f4} corresponds to contractions of the potential of an initially isotropic oscillator at different rates. In this case, an approach towards the classical behavior is observed. The classical scaling function has a jump discontinuity at $w=0$, and its derivative shows a discontinuity at $w/E = \min_i \h_i$.

When both principal directions are contracted at the same rate, the classical work pdf is uniform on $(0,\h_1)$. Because in this case transitions between states with even and odd energy quantum numbers are forbidden the quantum work pdf does not approach the classical case as displayed in figure~\ref{f6}.

Figure~\ref{f2} illustrates the limiting case, in which the potential is suddenly contracted in one direction while it remains unchanged in the orthogonal direction. With increasing initial energy $E$ the classical behavior which displays a square-root singularity at $w=0$ is approached. 

Finally, in figure~\ref{f8} the potential is expanded in one and contracted in the other direction. In this case, the oscillator may both perform and absorb work. For large values of the initial energy the classical behavior is approached with its characteristic logarithmic singularity at $w=0$.

In all cases, at finite energies, also scaled work values which lie outside the support of the classical work pdf and, hence, are forbidden in the classical case occur with finite probability.

\section{Conclusions}
Starting with a  comparison of different formulations of the density matrix of a microcanonical
state, we determined the work statistics of a closed system initially staying in a microcanonical state upon a change of control parameters according to a prescribed protocol  $\La =\{\l(t)|t_0 \leq t \leq t_f \}$  and presented  microcanonical fluctuation theorems in terms of the work pdf and the work probability, (\ref{FT}) and (\ref{ftqq}), respectively. These relations follow from a detailed balance-like relation for the transition probabilities (\ref{db}), which generalizes the detailed balance symmetry of autonomous systems to non-autonomous cases, and which also gives rise to a Jarzynski-type equation (\ref{dndm}). This microcanonical  Jarzynski relation allows one to express the degrees of degeneracy  of the Hamiltonian reached at the end of the protocol in terms of those of the initial Hamiltonian and to determine the microcanonical entropy of a system with the final Hamiltonian.

Upon averaging both sides of the microcanonical Jarzynski equality with  Boltzmann weights corresponding to the finally reached energies the standard,  canonical Jarzynski equality is recovered. Similarly, the canonical  Tasaki-Crooks relation is obtained from the fluctuation theorems by means of a canonical average.

In the case of a microcanonical initial state no simple formulation of the Tasaki-Crooks relation in terms of the characteristic function of work could be obtained apart from an integral equation connecting the characteristic functions for the forward and the backward protocol in a complicated way.
We refrained from giving this relation here.

As an example we determined the work statistics for sudden deformations of a two dimensional oscillator with respect to its principal directions.
Since for a one-dimensional oscillator any deformation protocol of finite duration can be mapped onto an equivalent sudden potential curvature change \cite{DL}  the obtained results apply for all protocols which extend over a finite time span and describe  squeezing and dilatation of a two-dimensional oscillator with respect to its principal directions. We though did not study more general cases which may contain arbitrary rotations of the potential.

We found that for transitions between oscillators with rational frequency ratios before and after the deformation the work distributions typically approach the respective classical distributions for increasing initial energies. The classical distributions can be described by scaling functions of the relative work $w/E$. This scaling function only depends on the frequency ratios $\o^f_i/\o^0_i$ of the two principal axes before and after the deformation.
A different distribution than the classical one was found for uniform deformations of an isotropic oscillator because then transitions between states with even and odd energy quantum numbers are strictly forbidden. Only if one restricts the work statistics to those work values corresponding to allowed transitions, the approach to the classical limit is recovered.
\appendix
\section{Classical Work pdfs}
In the framework of classical Hamiltonian mechanics the work pdf for a sudden change of the principal frequencies of a two-dimensional harmonic oscillator initially prepared in a microcanonical state at energy $E$ becomes
\begin{eqnarray}
p^{\mbox{cl}}_\La(w|E) &= \O_E^{-1} \i d^2 \bp \i d^2 \bq \:\d \left (w-\frac{m}{2} (\D \o^2_1 q_1^2 +\D \o^2_2 q_2^2)\right ) \nonumber\\
&\quad \times \d \left ( E-\frac{\bp^2}{2m} - \frac{m}{2} \left ((\o^{(0)}_1)^2 q_1^2 + (\o^{(0)}_2)^2 q_2^2 \right) \right )\:,
\label{pcla}
\end{eqnarray}
where
\begin{eqnarray}
\O_E &= \i d^2 \bp \i d^2 \bq \:\d \left ( E-\frac{\bp^2}{2m} - \frac{m}{2} \left ((\o^{(0)}_1)^2 q_1^2 + (\o^{(0)}_2)^2_2 q_2^2 \right ) \right ) \nonumber \\
&= (2 \p)^2 \frac{E}{\o^{(0)}_1 \o^{(0)}_2}
\label{O}
\end{eqnarray}
denotes the density of states of the initial microcanonical state and
$\D \o^2_i = (\o^{(f)}_i)^2 -(\o^{(0)}_i)^2$ quantifies the change of the frequency of the $i$th principal component. Here the integrals are extended over the phase space. The first delta-function under the integral on the right hand side of (\ref{pcla}) collects those initial phase space points which give the work $w$ and the second delta-function specifies the microcanonical state at energy $E$.  Carrying out the momentum integration and introducing the dimensionless coordinates $x_i = q_i/l_i$ one obtains for the classical work pdf the scaling relation
\be
p^{\mbox{cl}}_\La(w|E) = \r(w/E)/E\:,
\ee{pr}
where the scaling function $\r(z)$ becomes
\be
\r(z) = \frac{1}{\p} \i_G d^2 \bx \: \d(z - \h_1 x_1^2 -\h_2 x^2_2)\:.
\ee{r}
The integration is extended over the unit disk $G=\{\bx|1>x^2_1+x^2_2\}$. The dimensionless parameters $\h_i =\D \o^2_i/(\o^{(0)}_i)^2$ are measures of the potential deformation. The support of the scaling function is determined by those values of the scaled work variable $z$ for which a part of the conic section $z=\h_1 x^2_1 + \h_2 x^2_2$ lies within the unit disk $G$.
Depending on the signs of $\h_1$ and $\h_2$ one finds the following explicit results for the scaling function $\r(z)$:
\be
\fl
\r(z) = \frac{2}{\p \sqrt{|\h_1 \h_2|}} \left \{
\begin{array}{ll}
\displaystyle \frac{\p}{2} \Th(\frac{z}{\h_1}) \Th(1-\frac{z}{\h_1}) \\*[3mm]
\displaystyle +\Th(\frac{z}{\h_1}-1) \Th(1-\frac{z}{\h_2})\\*[3mm]
\displaystyle \times \arcsin\sqrt{\frac{(\h_2-z) \h_1}{(\h_2-\h_1)z}} & \mbox{for}\; \h_1 \h_2 >0,\; |\h_1| < |\h_2|\\
&\\
\displaystyle \Th(z) \Th(1-\frac{z}{\h_2}) \mbox{Arsinh} \sqrt{\frac{(z-\h_2)\h_1}{(\h_2-\h_1) z}}\\*[4mm]
\displaystyle + \Th(-z) \Th(1-\frac{z}{\h_1}) \mbox{Arsinh} \sqrt{\frac{(\h_1-z)\h_2}{(\h_2-\h_1) z}} \quad & \mbox{for}\; \h_1 <0,\;\h_2>0\: .
\end{array}
\right .
\ee{r2}
The upper form of $\r(z)$ applies if the oscillator undergoes a compression or an expansion in both principal directions provided $\h_1< \h_2$; the opposite case, $\h_1> \h_2$, follows by exchanging $\h_1$ and $\h_2$ with each other. The second form applies for an expansion in the 1-direction and a compression in the 2-direction.
In the remaining limiting cases one obtains
\be
\fl
\r(z)= \frac{1}{|\h|} \Th(\frac{z}{\h}) \Th(1-\frac{z}{\h}) \left \{
\begin{array}{ll}
\displaystyle 1&\mbox{for} \; \h_1=\h_2\equiv\h\\
\displaystyle \frac{2}{\p} \sqrt{\frac{\h}{z}-1}\quad&\mbox{for} \;\h_1=0,\;\h_2\equiv \h \; \mbox{or} \; \h_2=0,\;\h_1\equiv \h\:.
\end{array}
\right . 
\ee{r3}
Finally, the classical work pdf of a one-dimensional harmonic oscillator with a microcanonical initial state scales as
\be
p^{\mbox{cl, 1d}}_\La(w,E) = \r^{\mbox{1d}}(w/(\h E)) /( \h E)\:,
\ee{pcl1}
where
\be
\r^{\mbox{1d}}(z) = \frac{\Th(z) \Th(1-z)}{\p \sqrt{z-z^2}}\:,
\ee{rcl1}
with the deformation measure $\h=(\o^f/\o^0)^2-1$. Note that this pdf has square-root singularities at $w=0$ and $w=\h E$ whereas the two-dimensional case leads only to the singularity at $w=0$. \\


\begin{thebibliography}{99}
\bibitem{G} Gibbs WJ 1902 {\it Elementary Principles of Statistical Mechanics} (Scribner's Sons, New York)
\bibitem{Gibbs} Gibbs JW 1960 {\it Elementary Principles in Satistical Mechanics} (Dover, New York, reprint of the 1902 edition)
\bibitem{M} M\"unster A 1969 {\it Statistical Thermodynamics} Vol I (Springer Verlag, Berlin)
\bibitem{K} Khinchin AI 1949 {\it Mathematical Foundations of Statistical Mechanics} (Dover, New York)
 \bibitem{Campisi} Campisi M 2007 {\it Physica A} {\bf 385} 501
\bibitem{BMR} Barre J, Mukamel D and Ruffo S 2001 {\it Phys. Rev. Lett.} {\bf 87} 030601
\bibitem{CK} Campisi M and Kobe DH 2010 {\it Am. J. Phys.} {\bf 78} 608
\bibitem{Gross} Votyakov EV, Hidmi HI, De Martino A and Gross DHE 2002 {\it Phys. Rev. Lett.} {\bf 89} 031101
\bibitem{Ph} Phillips WD 1998 {\it Rev. Mod. Phys.} {\bf 70} 721
\bibitem{BDZ} Bloch I, Dalibard J and Zwerger W 2008 {\it Rev. Mod. Phys} {\bf 80} 885
\bibitem{CTH} Campisi M, Talkner P and H\"anggi P 2011 {\it Rev. Mod. Phys.} {\bf 83}, 771; ibid 2011 {\it Rev. Mod. Phys.} {\bf 83} 1653
\bibitem{CBK} Cleuren B, Van den Broeck C and Kawai R 2006 {\it Phys. Rev. Lett.} {\bf 96}, 050601
\bibitem{THM} Talkner P, H\"anggi P and Morillo M 2008 {\it Phys. Rev. E} {\bf 77}, 051131
\bibitem{CPRE} Campisi M 2008 {\it Phys. Rev. E} {\bf 78}, 012102
\bibitem{H} Holevo AS 2001 {\it Statistical Structure of Quantum Theory} (Springer, Berlin)
\bibitem{J} Jarzynski C 1997 {\it Phys. Rev. Lett.} {\bf 78} 2690
\bibitem{C} Crooks GE 1999  {\it Phys. Rev. E} {\bf 60} 2721
\bibitem{Tasaki} Tasaki H 2000 {\it arXiv:0009244 [cond-mat]}
\bibitem{TH} Talkner P and H\"anggi P 2007 {\it J. Phys. A} {\bf 40} F569
\bibitem{YKT} Yi J, Kim YW and Talkner P 2012 {\it Phys. Rev. E} {\bf 85} 051107
\bibitem{TBH} Talkner P, Burada PS and H\"anggi P 2008 {\it Phys. Rev. E} {\bf 78} 011115
\bibitem{DL} Deffner S and Lutz E, 2008 {\it Phys. Rev. E} {\bf 77} 021128
\bibitem{BR} O Bratteli and DW Robinson 1997 {\it Operator Algebras and Quantum Statistical Mechanics} 2, p. 366 (Springer, Berlin)
\bibitem{DS} Dunford N and Schwartz JT 1958 {\it Linear Operators, Part I: General Theory} (John Wiley and Sons, New York)
\bibitem{AG} Andrieux D and Gaspard P (2008) {\it Phys. Rev. Lett.} {\bf 100}, 230404
\bibitem{CTH2011} Campisi M, Talkner P and H\"anggi P 2011 {\it Phil. Trans. R. Soc. A} {\bf 369} 291
\bibitem{Hertz} Hertz P 1910 {\it Ann. Phys. } {\bf 33} 537
\bibitem {Dunkel} Dunkel J and Hilbert S 2013 {\it arXiv:1304.2066 [cond-mat.stat-mech]}
\bibitem{TLH} Talkner P, Lutz E and H\"anggi P 2007 {\it Phys. Rev. E} {\bf 75}, 050102.
\bibitem{GR} Gradstheyn IS and Ryzhik IM 1965 {\it Table of Integrals, Series and Products} (Academic Press, New York)
\bibitem{DAL} Deffner S, Abah O and Lutz E 2010 {\it Chem. Phys.} {\bf 375} 200
\end{thebibliography}
\end{document}